%% file: asme2e.tex
\documentclass[twocolumn,10pt]{asme2e}

\usepackage{amsmath}
\usepackage{graphicx}
\usepackage{subfigure}

\title{DEPOSITION OF SAND PARTICLES ON A SOLID SUBSTRATE IN A HIGH-TEMPERATURE SUBSONIC FLOW}

\author{Rahul Babu Koneru\thanks{Address all correspondence to this author at rkoneru[at]umd.edu.}
    \affiliation{
	Department of Aerospace Engineering\\
        University of Maryland\\
	College Park, Maryland 20742\\
    }	
}

\author{Luis Bravo \\
       {\tensfb Muthuvel Murugan }\\     
       {\tensfb Anindya Ghoshal }     
    \affiliation{DEVCOM US Army Research Laboratory\\
	Aberdeen Proving Ground, Maryland, 21005\\
    }
}
\author{Alison Flatau
    \affiliation{
	Department of Aerospace Engineering\\
        University of Maryland\\
	College Park, Maryland 20742\\
    }	
}

\input{dfn}

\begin{document}

\maketitle    

\begin{abstract}
{\it Ingestion of sand particles into gas turbine engines has been observed to cause damage to engine components and in
some cases leads to catastrophic failure. One such mechanism responsible for engine failure occurs through the deposition
of molten particles on the turbine blades in the hot-section of the engine. The deposited material reacts chemically and
penetrates the thermal barrier coating (TBC) on the turbines blades eventually damaging them. As high-efficiency engines
are being pursued, the peak operating temperatures are bound to increase well above the softening temperature of the
solid particulates. In this work, we investigate the deposition of sand particles on a solid substrate using two-way
coupled Euler-Lagrange simulations. In these simulations, hot gas at 1700 K is issued from a circular inlet at Mach
0.3. Simultaneously, spherical sand particles, modeled after the Calcia-Magnesia-Alumino-Silicates(CMAS), are injected at
a constant mass flow rate of 1 gram per minute. The deposition of these particles on a solid substrate, placed 20 cm away
from the inlet along the axial direction, is investigated. These simulations are modeled after the experiments performed
using the Hot-Particle Ingestion Rig (HPIR) facility at the Army Research Laboratory. The particle rebound and
deposition model given by Bons et al. [``A simple physics-based model for particle rebound and deposition in
turbomachinery.'' Journal of Turbomachinery 139.8 (2017)] is implemented and used in this work. Euler-Lagrange
simulations are carried out for three different synthetic sand particles (CMAS, AFRL 02 and AFRL 03). To isolate the
effects of Stokes number mono-dispersed particles are injected with a Gaussian spatial distribution. The effect of
material properties and particle size on particle properties such as number of particle depositions, rebound velocity
and coefficient of restitution are investigated.}
\end{abstract}

\begin{nomenclature}
\entry{TBC}{Thermal Barrier Coating.}
\entry{EBC}{Environmental Barrier Coating.}
\entry{YSZ}{Yttria stabilized Zirconia.}
\entry{CMAS}{Calcia-Magnesia-Alumino-Silicate.}
\entry{AFRL}{Air Force Research Laboratory.}
\entry{HPIR}{Hot Particulate Ingestion Rig.}
\entry{$(.)_p$}{Particle property.}
\entry{$(.)_s$}{Surface/coupon property.}
\entry{$V_{n1}$}{Impact velocity in the normal direction.}
\entry{$V_{n2}$}{Rebound velocity in the normal direction.}
\entry{$V_{n1crit}$}{Critical velocity of impact in the normal direction.}
\entry{$\nu$}{Poisson's ratio.}
\entry{$E_c$}{Composite Young's modulus.}
\end{nomenclature}


\section*{INTRODUCTION}
\begin{figure*}[t]
\begin{center}
  \includegraphics[scale=0.40]{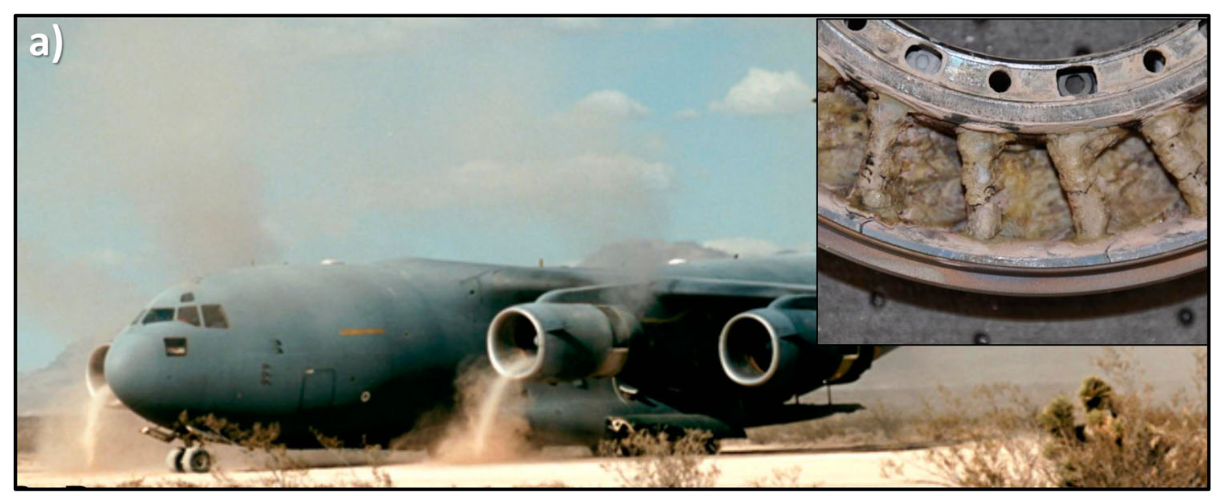}
\end{center}
\caption{SAND INGESTION DURING TAKE-OFF AND MOLTEN PARTICULATE DEPOSIT (INSET).}
\label{fig:c17} 
\end{figure*}

Rotorcrafts operating in dusty environments have been observed to suffer structural damage due to the ingestion of solid
particulates into the gas turbine engines (GTEs). The ingested particles can cause erosion due to repeated impact,
accumulate in air pathways leading to blockages and cause material degradation due to molten particulate deposits on the
hot sections of the GTE. An aircraft undergoing sand ingestion is shown in Fig. \ref{fig:c17} along with molten
particulate deposits on the engine vanes in the inset. 
Environmental barrier coatings (EBC) offer protection against
kinetic impacts while the inertial particle separators filter out larger particles ($>\ 75 \mu m$) without a significant
pressure drop at the compressor inlet. Smaller particles, on the other hand, pass through the cold section, melt in the
combustion chamber and the resulting molten material, comprised of calcia-magnesia-alumina-silicates (CMAS), adheres to
and damages the thermal barrier coating on the hot-section components. The molten CMAS material has been observed to
infiltrate, react chemically with the thermal barrier coating (TBC) and solidify into a glassy coating as it cools down
\cite{nieto2020calcia} \cite{murugan2021search}. Apart from structural damage, the CMAS penetration has also been
observed to alter the thermal properties of the TBCs such as volumetric heat capacity and thermal conductivity
\cite{nieto2020calcia}. Some of the mitigation strategies involve tailoring the TBC microstructure \cite{kang2018high}
and accelerating the chemical reaction time between the molten CMAS deposit and the coating to induce solidification and
thus prevent penetration. One popular choice for the TBC on hot-section components is a mixture of Yttria
($Y_2O_3$)-stabilized Zirconia ($ZrO_2$) (YSZ). This particular coating has several desirable features (for
high-temperature application) such as high melting point ($\sim 2700\ ^oC$), low thermal conductivity at elevated
temperatures ($\sim 2.3\ W\cdot m^{-1}\cdot K^{-1}$ at $1000\ ^oC$) and high thermal-expansion coefficient ($\sim 11
\times 10^{-6}\ ^oC^{-1}$) \cite{padture2002thermal}. With the performance envelope of the GTEs ever expanding, the
operating temperatures are bound to increase which will only exacerbate CMAS attack. At the US Army Research Laboratory
(ARL), research is underway to develop novel `sandphobic' TBCs for high-temperature applications. \par
Recent progress in sand modeling 
at ARL employed a multiscale approach to investigate particle transport, collisions, and surface impact. This involved investigation of the
interactions of binary fluid droplets, resembling molten sand particles, using the volume-of-fluid approach
\cite{ganti2020binary,ganti2020binary2}, smoothed particle hydrodynamics to investigate high-velocity impacts of molten
sand particles \cite{chaussonnet2020smoothed} and two-way coupled Euler-Lagrange (EL) simulations to investigate the
effect of Stokes number on particle impact and heat flux on a transonic guide vane \cite{bravo2020physical,bravo2021uncertainty}. 
Building upon this past work at ARL, high-fidelity computational fluid dynamics analysis is carried out to investigate
the underlying transport and deposition physics using two-way coupled EL simulations in a gas turbine engine relevant environment. \par  
The interaction between the particles and the (TBC) surfaces is crucial for particle deposition. The loss of energy of
an impacting particle can be from a loss of kinetic energy and from energy required to overcome adhesive forces. There
are multiple sources for the loss of energy from  adhesion such as van der Walls force, capillary force, gravitational
force and electrostatic force. A survey of the existing literature, points to various deposition models. In the critical
velocity model of Brach \& Dunn \cite{brach1992mathematical}, the particle deposition is determined by the critical
velocity of the impacting particle. The critical velocity is a function of material properties such as Young's modulus,
yield stress and particle density. Below this critical velocity, the particle expends all its kinetic energy and
deposits on the surface. This model was developed for low-velocity impacts. Another model proposed by Sreedharan \&
Tafti \cite{sreedharan2011composition} is based on the sticking probability which is computed based on the critical
viscosity approach. The critical viscosity is based on the softening temperature of a particle above which the viscosity
decreases rapidly. The particle deposits if the actual sticking probability is greater than a probability drawn randomly
from a uniform distribution. Additionally, the particle also deposits if the particle temperature is above the softening
temperature. Drawing on these models, Singh \& Tafti \cite{singh2015particle} proposed a hybrid model combining the
critical velocity and critical viscosity approaches. In the current work, the deposition model developed by Bons \et\
\cite{bons2017simple} is implemented and used. This model depends on the elastic-plastic deformation of a spherical
particle idealized as a cylinder when treating the collision with the surface. The nature of the deformation, elastic or
plastic, is determined based on the critical velocity of the particle. Based on this information, a contact area between
the particle and the surface is computed which in turn is used to compute the work done by the adhesive forces. The
kinetic energy at impact has to be greater the work of adhesion for the particle to rebound or else the particle
deposits on the surface. Finally, the rebound normal and tangential velocities are computed using a soft-sphere analogy.
In the current work, deposition characteristics of three different sand particles are investigated on a material coupon.
The different sand particles used are CMAS, AFRL 02 and AFRL 03 particles which are treated as rigid spheres. The
simulations are modeled after the experiments performed at the US Army Research Laboratory using the hot-particulate
ingestion rig (HPIR) \cite{nieto2020calcia}\cite{ghoshal2018molten}. 

\section*{GOVERNING EQUATIONS} \label{sec:geq}

In this work, an Euler-Lagrange framework is used to handle the multiphase system. For the gas phase, compressible
Navier-Stokes equations coupled with hydrodynamic drag and interphase heat-transfer are solved. These are given below.
\begin{align}
  &\pp{\rho}{t} + \div \left( \rho \ug \right) = 0 \label{eq:gas_mass}\\
  &\pp{\rho \ug}{t} + \div \left( \rho \ug \otimes \ug \right) + \nabla p = \div \taub - \fqs \label{eq:gas_mom}\\
  &\pp{\rho E}{t} + \div \left( \left( \rho E + p \right) \ug \right) - \div \left( \kappa \nabla T \right) = \div
  \left( \ug \cdot \taub \right)- \gqs - q \label{eq:gas_en}
\end{align}
where the shear stress $\taub$ and the shear-rate tensor $\mathbf{S}$ are defined as,
\begin{align}
  &\taub = 2 \mu \mathbf{S} \label{eq:tau}\\
  &\mathbf{S}  = \frac{1}{2} \left( \nabla \ug + \nabla \ug^T \right) - \frac{1}{3}\div \ug. \label{eq:S}
\end{align}
The dynamic viscosity ($\mu$) of the flow, computed using the Sutherland's law, is given by,
\begin{equation}
  \mu(T) = \frac{1.485 \times 10^{-6} T^{1.5}}{T+110.4}
  \label{eq:mu}
\end{equation}
and the coefficient of thermal conduction ($\kappa$) is given by,
\begin{equation}
  \kappa(T) = C_{p,gas} \kappa_0 T^{n}
  \label{eq:kappa}
\end{equation}
where, $\kappa_0= 6.25 \times 10^{-7} kg\cdot \left( s \cdot m \cdot K^n \right)^{-1}$ and $n=0.7$.  The hydrodynamic
force coupled back to the gas is represented by $\fqs$ and similarly the contributions from the interphase work done and
heat transfer is represented by $\gqs$ and $q$ respectively.  \\ 
The system of equations for Lagrangian point-particles are:
\begin{align}
  &\dd{\xp}{t} = \vp \label{eq:pcl_pos}\\
  &\dd{\vp}{t} = \Fqs \label{eq:pcl_vel}\\
  &\dd{\Tp}{t} = Q \label{eq:pcl_en}
\end{align}
where the quasi-steady force ($\Fqs$) and the heat-transfer between the phases $Q$ are given by,
\begin{align}
  &\Fqs = \frac{\ug-\vp}{\tau_v} \label{eq:Fqs}\\
  &  Q  = \frac{T - \Tp}{\tau_T}. \label{eq:Qht}
\end{align}
The hydrodynamic and the thermal time scales are represented by $\tau_v$ and $\tau_T$ respectively.
\begin{align}
  &\tau_v = \frac{4}{3} \frac{\rho_g}{\rp} \frac{d_p}{C_D} \frac{1}{|\vp-\ug|} \label{eq:tau_v}\\
  &\tau_T = \frac{1}{6} d_{p}^{2} C_{p,particle} \frac{\rho_p}{\kappa Nu} \label{eq:tau_T}
\end{align}
The drag coefficient $C_D$ proposed by Loth \cite{loth2008compressibility} is used while the Nusselt number ($Nu$) is
calculated using the correlation given by R. Ranz \& Marshall \cite{ranz1952evaporation}.\\
The system of equations are solved using the massively parallel code Athena-RFX
\cite{poludnenko2010naval,poludnenko2010interaction}. The integration is carried out using an unsplit corner transport
upwind (CTU) algorithm \cite{gardiner2005unsplit} based on the work of Collela \cite{colella1990multidimensional}. An
approximate HLLC Riemann solver is used along with a piecewise-parabolic method (PPM) for flux reconstruction resulting
in a spatial accuracy of $3^{rd}$-order.  The particle phase equations, Eqs. \ref{eq:pcl_pos}-\ref{eq:pcl_en}, are
integrated in time using a semi-implicit predictor-corrector algorithm. This offers better numerical stability and the
errors in particle trajectories remain bounded in time. This property of preserving particle trajectories over long
integration times is associated with symplectic integrators typically used for Hamiltonian systems
\cite{kozak2019novel}. The Eulerian fluid properties are interpolated to the particle location using a 5-point stencil
WENO scheme which is $5^{th}$-order accurate for smooth flows \cite{kozak2020weno}.

\subsection*{Model for Particle Deposition}

The particle deposition model proposed by Bons \et \cite{bons2017simple} is implemented and used in this work to explore
the particle deposition mechanism. In this model, the spherical particle is idealised as a cylinder with an equal volume
impacting the substrate along the axis of rotation of the cylinder. The particle-surface collision is treated as a
spring-damper system and the effect of surface inhomogeneities are accounted for via an adhesion model. In this model,
the quantity that determines if a particle is deposited is the normal coefficient of restitution \corn\ which is defined
as the ratio of the rebound normal velocity ($V_{n2}$) and the impact normal velocity ($V_{n1}$). The particle deposits
on the surface when \corn\ is equal to zero and for all the other values the particle rebounds off the surface. The
procedure for computing $V_{n2}$ is laid out in the next paragraph. 

As mentioned in the preceding paragraph, the particles are represented as cylinders with the same volume as that of the
sphere. The resulting length of the cylinder $l=2 d_p/3$. Based on the material properties of the particle such as
yield stress ($\sigma_y$), particle density ($\rho_p$) and Young's modulus ($E_c$), the critical normal velocity
$V_{n1crit}=\sigma_y/(\rho_p E_c)^{0.5}$ is computed at which plastic deformation of the particle begins. In this case, a
composite modulus is defined based on the Poisson's ratio ($\nu$) and the Young's moduli of the particle and the surface. 
\begin{equation}
  \frac{1}{E_c} = \frac{1-\nu_p^2}{E_p} + \frac{1-\nu_{s}^2}{E_{s}}
  \label{eq:Youngs}
\end{equation}
Furthermore, the impacting particle expends additional energy to overcome the adhesive forces. The resulting kinetic
energy of the particle is computed by subtracting the work of adhesion ($Wa$) from the critical elastic energy ($E_{crit}$). The work of
adhesion depends on two quantities 1) contact surface area ($A_{cont}$) and 2) deformation of the particle ($w$) upon
impact. The work of adhesion is computed using the expression $W_a=A_{cont}\gamma_{s}$ where $\gamma_{s}$ is the surface
free energy taken to be 0.8. The contact area is computed using the following expression
\begin{equation}
  A_{cont} = A_{crit}\left[ a+b \left(\frac{w_{max}}{w_{crit}} \right)^{c} \right]
  \label{eq:Acont}
\end{equation}
where $a=0.1$, $b=1/7$, $c=0.5$ are empirical constants and $w_{max}$, $w_{crit}$ denote the maximum plastic deformation
and maximum elastic deformation respectively. The surface area at maximum elastic deformation is given by
\begin{equation}
  A_{crit} = \frac{\pi d_p^2 l}{4(l-w_{crit})}.
  \label{eq:Acrit}
\end{equation}
For the definitions of $w_{max}$, $w_{crit}$ and $E_{crit}$ see \cite{bons2017simple}.

Upon collision, some particles lose all the elastic energy to the work of adhesion and deposit on the surface. While
particles with enough elastic energy undergo an elastic rebound. The normal and tangential components of the rebound
velocity are given below. 
\begin{align}
  &V_{n2} = -CoR_n V_{n1} \\
  &V_{t2} = V_{t1} - \beta V_{n1} \cos^2(\alpha) (1+\corni)\left( 1+ \frac{2W_a}{\corni} m_p V_{n1}^2 \right)
  \label{eq:rebound}
\end{align}
The ideal normal coefficient of restitution (\corni) is the ratio of $V_{n1crit}$ and $V_{n1}$. The angle of impact is
calculated using $\alpha = \arctan({V_{n1}/V_{t1}})$ where $V_{t1}$ is the impact tangential velocity.  Finally, the
impulse ratio $\beta$ is given by 
\begin{equation}
  \beta = \frac{V_{t1}}{V_{t2}} \left[ 1 + \corni^2 + \frac{2W_a}{m_p V_{n1}^2} \right]^{-0.5}.
  \label{eq:beta}
\end{equation}

\section*{SIMULATION SETUP}
\begin{figure}[t]
\centering
  \begin{subfigure}
    \centering
  \includegraphics[scale=0.25]{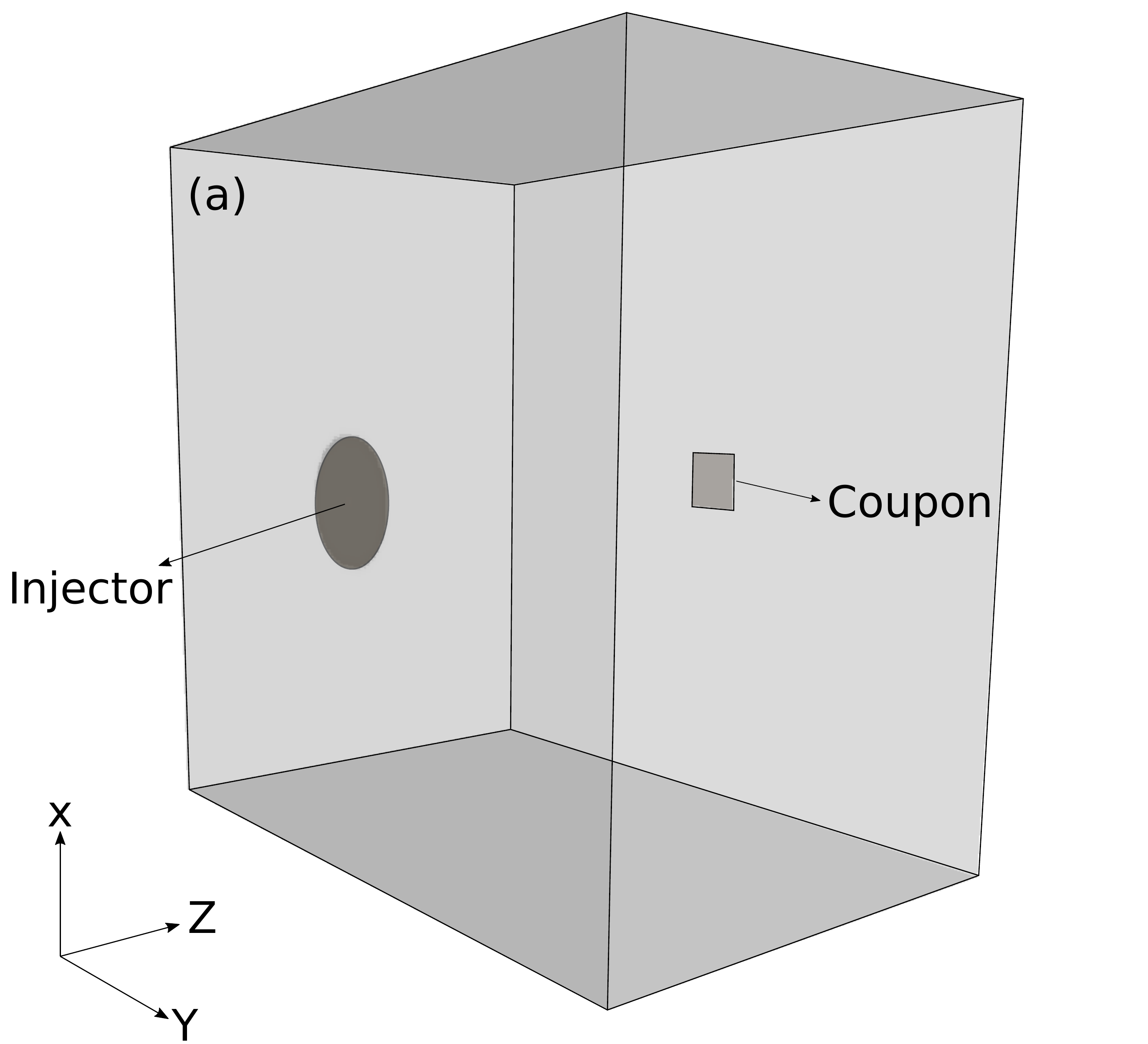}
  \end{subfigure}
  \begin{subfigure}
    \centering
  \includegraphics[scale=0.40]{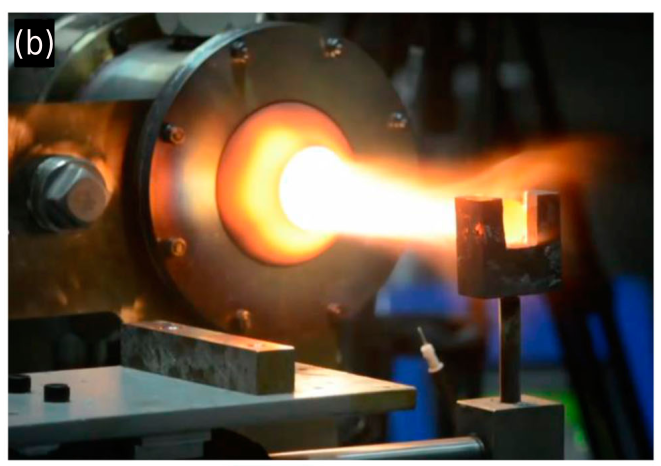}
  \end{subfigure}
  \caption{(a) SIMULATION SETUP AND (b) HPIR FACILITY \cite{nieto2020calcia}.}
\label{fig:setup} 
\end{figure}

The simulations are performed in a three dimensional (3D) Cartesian box with the jet aligned along the z-axis. A
schematic of the setup is shown in Fig. \ref{fig:setup}. Along the streamwise direction, the jet is issued from a
circular inflow region at one end and at the other end is the coupon. Apart from the inflow boundary condition, rest of
the boundaries are set to no-slip walls for the gas flow. For the particles on the other hand, the deposition boundary
condition is specified on the coupon. The grid spacing in the streamwise direction is 781.25 $\mu m$ and about 1.17 mm in the
other two directions which amounts to around 16.7 million cells in total.
The initial conditions of the gas and the particles are given in Table \ref{tab:ic} and are based on experimental measurements. The initial temperature and the
velocity of all the particles in the radial and axial directions is the same as the hot jet and the mass flow rate of
the particles is set to 1 gram/min. To maintain a steady mass flow rate, the particles are injected into the domain in pulses. Particles are currently assumed to be in thermal equilibrium. Thus, phase-change or surface chemistry is not modeled but will be addressed in future works. A photograph of the HPIR experimental facility is shown in Fig. \ref{fig:setup}(b).
\begin{table}[t]
\caption{INITIAL CONDITIONS}
\begin{center}
\label{tab:ic}
\begin{tabular}{c l l}
& & \\ 
\hline
Parameter & Value & Definition \\
\hline
$u_z$ & 250 m/s & Axial velocity of the jet\\
$u_r$ & 25  m/s & Radial velocity of the jet \\
T     & 1700  K & Temperature of the jet \\
d     & 0.05  m & Diameter of the jet \\
$l_{x,coup}$ & 0.254  m & Size of the coupon along the x-direction\\
$l_{y,coup}$ & 0.254  m & Size of the coupon along the y-direction\\
\hline
\end{tabular}
\end{center}
\end{table}
The ambient pressure and temperature are set to 1 atm and 298 K. The gas in this simulation is treated as air with
$\gamma=1.4$ and $R=287 J \cdot (kg \cdot K )^{-1}$.

In this work, three different particles are used which are primarily composed of calcia-magnesia-alumino-silicates
(CMAS). These are referred to as CMAS, AFRL 02 and AFRL 03. The AFRL particles are assumed to have identical chemical
composition in this work and hence have the same mechanical properties. The only difference between the two different
AFRL particles is in the size of the particles. The AFRL 02 particles are smaller in size and are used for test-bed
analysis while the (comparatively) larger sized AFRL 03 particles are synthesized for engine-level testing
\cite{ghoshal2019governing}. The mechanical properties of the AFRL particles are taken from the works of Bojdo, N. and
Filippone, A. \cite{bojdo2019simple} and Whitaker, S. M. \cite{whitaker2017informing} while that of the CMAS particles
are taken from a NASA report by Bansal, N. P., and Choi, S. R. \cite{bansal2014properties} on desert sand and CMAS glass.
The coupon is representative of bulk Yttria stabilized Zirconia (YSZ). The material properties related to the coupon
are accounted for in the deposition model via the Young's modulus ($E_s$) and the Poisson's ratio ($\nu_s$). Following
the work of Nieto \et \cite{nieto2018layered}, these values are set to $E_s=220$ GPa and $\nu_s=0.3$. The material
properties of the CMAS and the AFRL particles are listed in Table \ref{tab:afrl_prop}. These bulk properties are
computed using volume fraction weighted summation of individual constituents. The solid particle characterization
measurements were conducted at ARL to statistically quantify the particle distributions. Following this, the particle
diameters chosen were 20.15 $\mu m$ for the CMAS particles and 17.86 $\mu m$ and 22.89 $\mu m$ for the AFRL 02 and
AFRL 03 particles respectively. Finally, the yield stress of the particles is computed as a function of temperature given by the
\begin{equation}
  \sigma_y(T) = 200 - 0.255 (T-1000) MPa.
  \label{eq:sigy}
\end{equation}
\begin{table}[t]
\caption{MATERIAL PROPERTIES OF CMAS AND AFRL 02/03 PARTICLES} 
\begin{center}
\label{tab:afrl_prop}
\begin{tabular}{c l l}
& & \\ 
\hline
Parameter & CMAS & AFRL 02/03 \\
\hline
$\rho_p$         & 2690 $kg \cdot m^{-3}$       & 2547 $kg \cdot m^{-3}$       \\
$E_p$            & 92.3 GPa                     & 73.8 GPa                     \\
$\nu_p$          & 0.300                        & 0.235                        \\
$C_{p,particle}$ & 800 $J \cdot kg^{-1} K^{-1}$ & 863 $J \cdot kg^{-1} K^{-1}$ \\
\hline
\end{tabular}
\end{center}
\end{table}
\section*{RESULTS}
\begin{figure}[t]
\centering
  \begin{subfigure}
    \centering
  \includegraphics[scale=0.25]{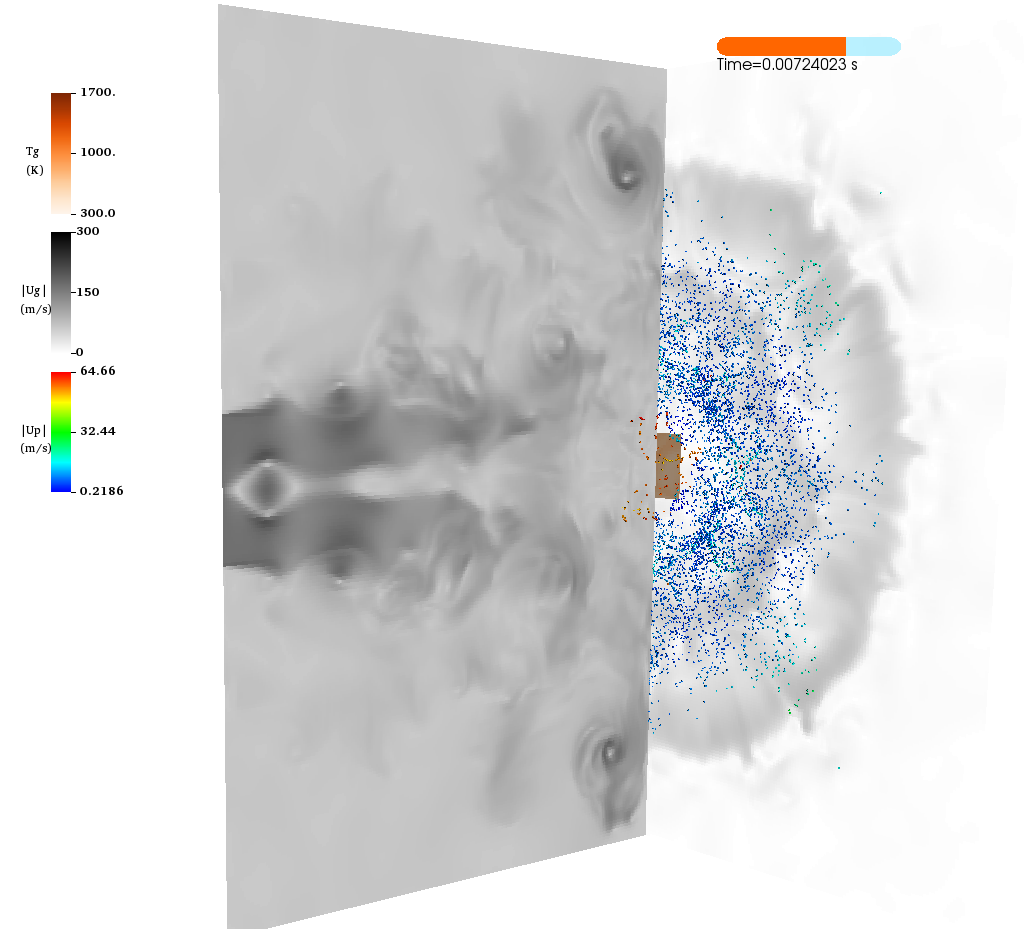}
  \end{subfigure}
  \begin{subfigure}
    \centering
    \includegraphics[scale=0.20]{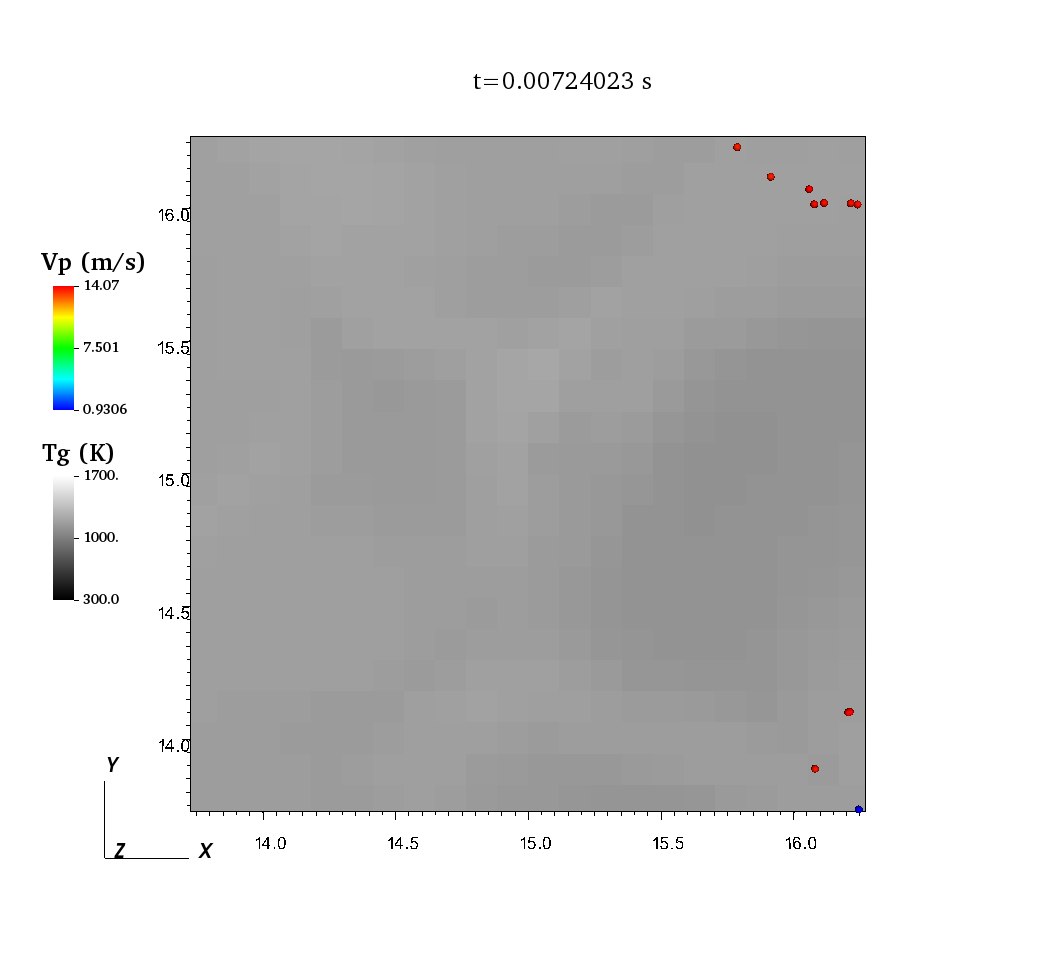}
  \end{subfigure}
  \caption{ (TOP) INSTANTANEOUS SNAPSHOT OF THE FLOW FIELD AND (BOTTOM) CLOSE-UP VIEW OF THE COUPON.}
\label{fig:snap} 
\end{figure}

The general feature of the flow can be seen in the instantaneous snapshot of the flow field in Fig. \ref{fig:snap}. The
vortex roll-up of the gas reminiscent of Kelvin-Helmholtz instability is seen as the hot gas enters into cooler and less
viscous ambient gas. Based on the injection parameters of the gas, the Mach number is 0.29 and the Reynolds number based
on the injector diameter is 44668. The location of the particles hitting the coupon can be seen in the
close-up image in Fig. \ref{fig:snap}

\begin{figure}[t]
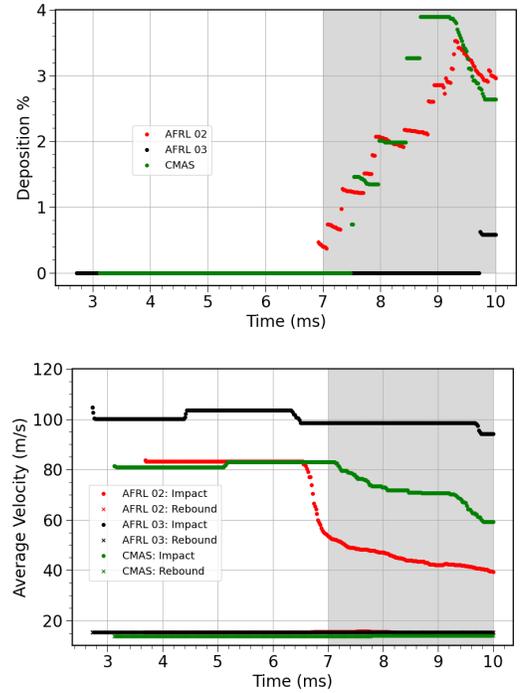

\centering
\begin{subfigure}
    \centering
    \includegraphics[scale=0.30]{images/afrl_cmas_dep_pct_shd}
  \end{subfigure}
  \begin{subfigure}
    \centering
    \includegraphics[scale=0.30]{images/afrl_cmas_Vimp_Vreb_sigY_temp_shd}
  \end{subfigure}
  \caption{PLOT OF THE CUMULATIVE PERCENTAGE OF PARTICLES BEING DEPOSITED ON THE COUPON (TOP) AND THE AVERAGE IMPACT AND REBOUND
  VELOCITIES OF THE PARTICLES (BOTTOM).}
\label{fig:dep_vel_reb} 
\end{figure}

\begin{figure}[t]
\begin{center}
  \includegraphics[scale=0.75]{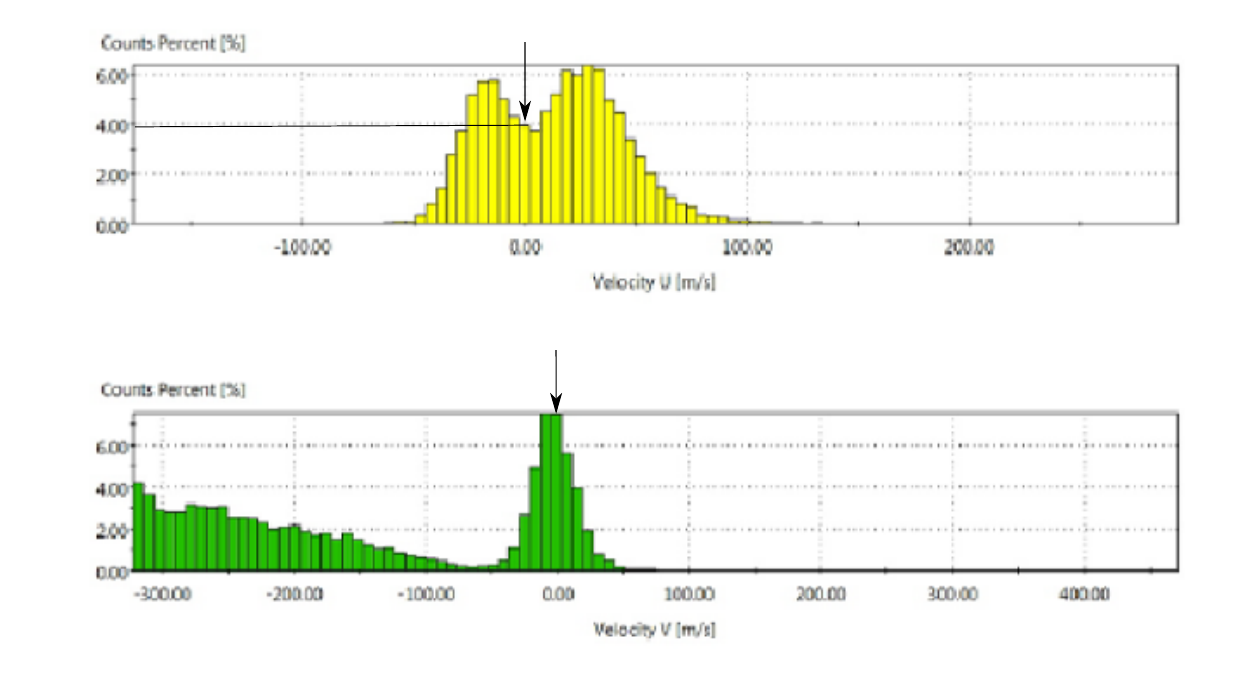}
\end{center}
\caption{REBOUND VELOCITIES OF SAND PARTICLES FROM THE HPIR EXPERIMENT. THE DEPOSITION PERCENTAGES ARE DENOTED BY THE
ARROWS.}
\label{fig:hpir_vel_reb} 
\end{figure}

In Fig. \ref{fig:dep_vel_reb}(a), the cumulative number of particles being deposited as a fraction of the total number of
particles hitting the coupon is represented as a percentage. In the bottom figure, average impact and
rebound velocity of all the particles hitting the coupon is shown. Among the three different particles, AFRL 03
particles show the least amount of deposition. CMAS and AFRL 02 on the other hand, have a peak deposition percentages of
about 4\% and 3.5\% respectively. A direct correlation can be seen between the average impact velocities of the
particles and the deposition percentage. The impact velocities of the AFRL 02 and the CMAS particles begin to dip at
around the same time particles begin depositing on the coupon. This is clearly an effect of the deposition model which
relies on the loss of kinetic energy during impact, in the normal direction, due to surface interaction to determine the
deposition rate. As the incoming particles lose their kinetic energy, the particles begin depositing on the coupon. The
average rebound velocity of all the particles is around 15 m/s. This suggests that the rebound velocity of the AFRL 03
particles is low but not quite low enough for the particles to deposit which explains the low deposition rate. The time
averaged distribution (in \%) of normal and tangential components of the particle rebound velocity from the experiment
are shown in Fig. \ref{fig:hpir_vel_reb}. The deposition percentage (where the velocity is 0) corresponding to the
normal velocity is about 4\% and that of the tangential component is a little over 7\%. While comparing with the
experiments it is important to note that the sand particles in the experiments were polydisperse in nature whereas the
particle sizes in the simulations are the volume averaged mean values of the polydisperse mix.
\subsection*{Effect of Stokes Number}
\begin{figure}[t]
\begin{center}
  \includegraphics[scale=0.30]{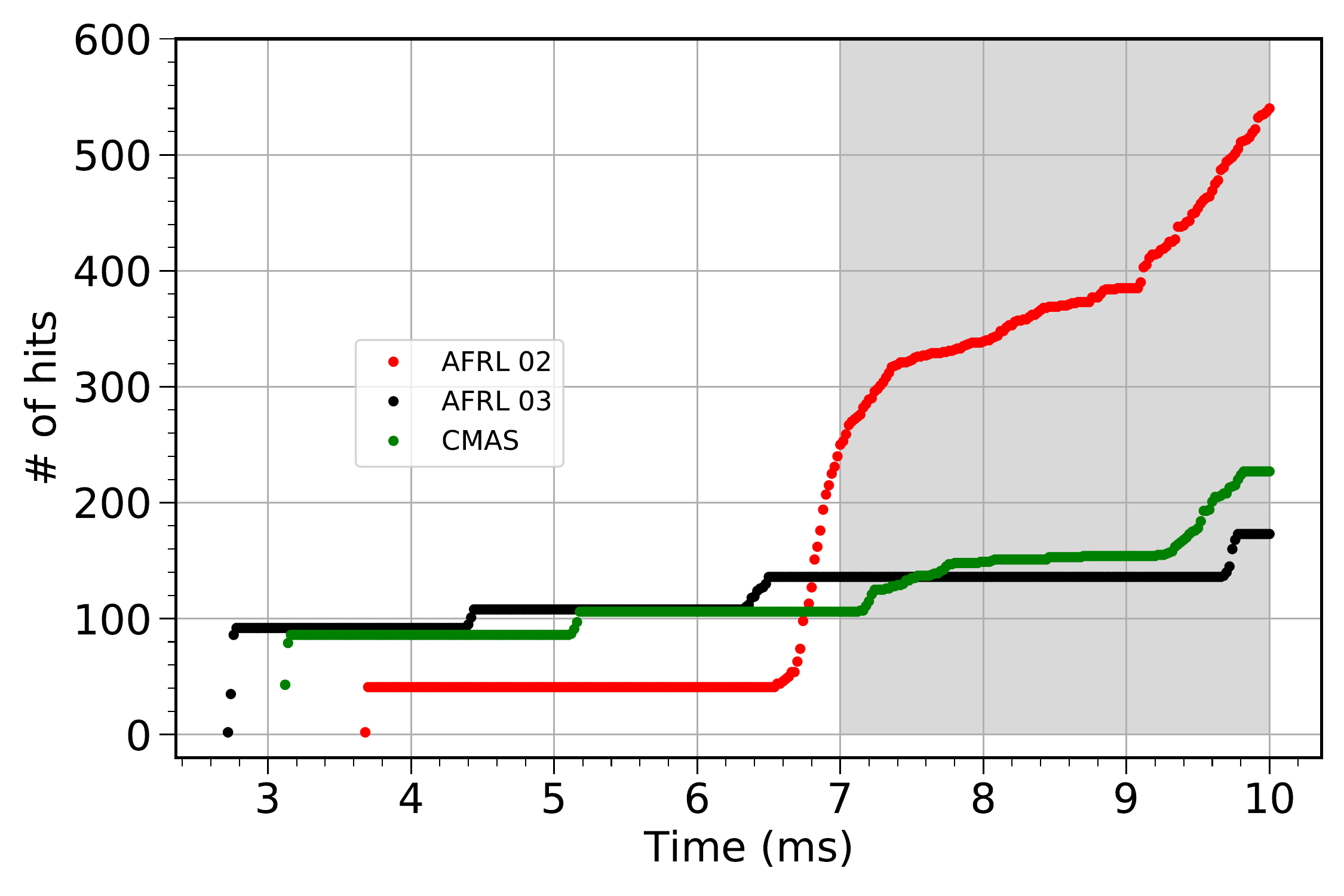}
\end{center}
\caption{CUMULATIVE SUM OF ALL THE PARTICLE IMPACTS ON THE COUPON.}
\label{fig:afrl_cmas_hits} 
\end{figure}

\begin{figure*}[t]
\centering
  \begin{subfigure}
    \centering
    \includegraphics[scale=0.25]{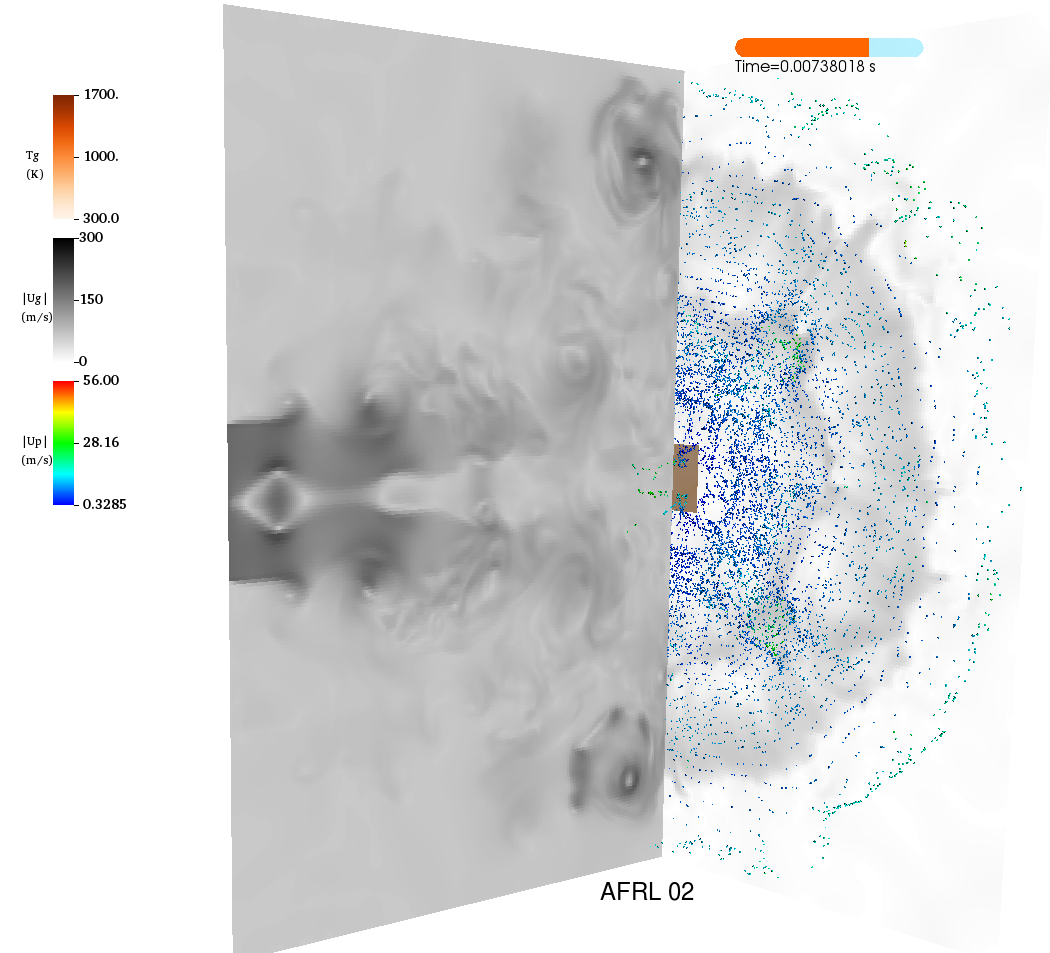}
  \end{subfigure}
  \begin{subfigure}
    \centering
    \includegraphics[scale=0.25]{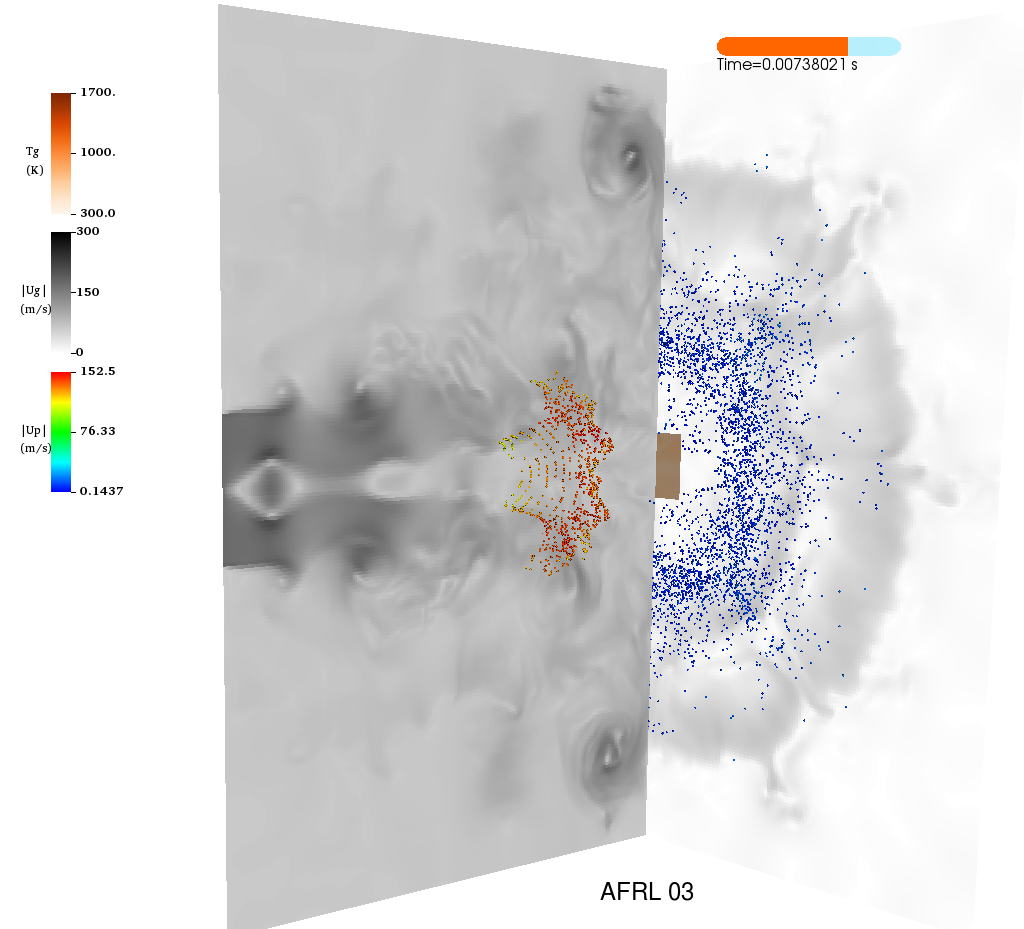}
  \end{subfigure}
  \begin{subfigure}
    \centering
    \includegraphics[scale=0.25]{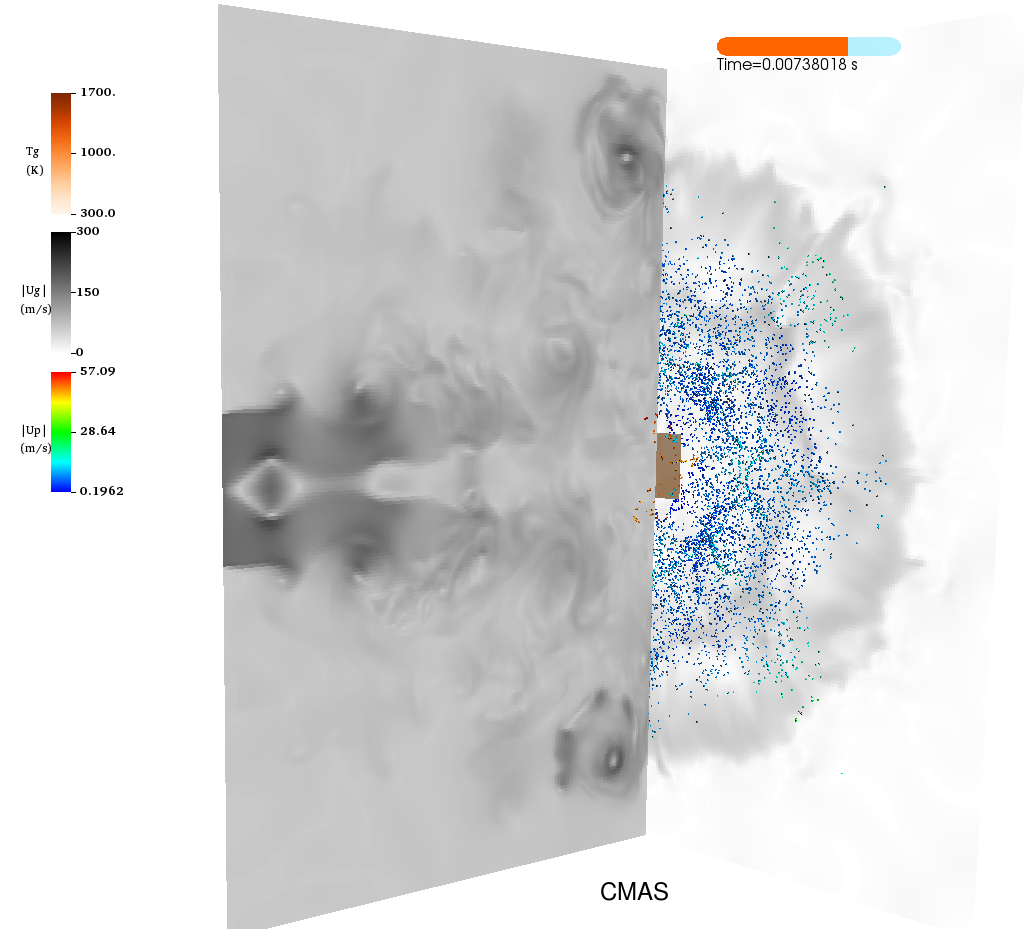}
  \end{subfigure}
  \caption{INSTANTANEOUS SNAPSHOTS OF THE FLOW FIELDS FOR THE THREE PARTICLES. AFRL 02 (TOP LEFT), AFRL 03 (TOP RIGHT)
  AND CMAS (BOTTOM).}
\label{fig:snap3} 
\end{figure*}

The AFRL 03 particles are about 30\% larger than the AFRL 02 particles and to investigate the effect of particle size it
is useful to compute the Stokes number of the particles. The Stokes number is computed using the formula
\begin{equation}
  Stk = \frac{\rho_p d_{p}^{2}}{18 \mu_f C_D} \cdot \frac{U_{inj}}{d_{inj}}
  \label{eq:stk}
\end{equation}
where $U_{inj}$ is the velocity magnitude of the gas injected into the domain. Based on this, the Stokes numbers come
out to 1.33, 2.56 and 1.93 for AFRL 02, AFRL 03 and CMAS particles respectively. Although all the particles are of the
same order of magnitude we see a difference in the deposition of the AFRL particles. It is likely that the larger AFRL
03 particles are more ballistic than the AFRL 02 particles and are hence less affected by the gas flow especially in the
vicinity of the coupon. This can be seen in the impact velocities of these particles in Fig. \ref{fig:dep_vel_reb} and
in the instantaneous snapshots of the flow fields, taken at about 7.4 ms, in Fig. \ref{fig:snap3}. In the case of the larger AFRL 03
particles, the trailing particles are about to reach the end of the domain whereas the AFRL 02 and the CMAS particles
can be seen already reaching the end of the domain and hitting the coupon. These trailing particles are a consequence of maintaining a uniform mass flow rate of particles wherein the particles are injected in pulses. The batch of AFRL 02 and CMAS particles accompanying the AFRL 03 particles has already reached the end of the domain. Further evidence of this argument can be
seen in the total number of particle impacts on the coupon shown in Fig. \ref{fig:afrl_cmas_hits}. The smaller AFRL 02
particles register the highest number of hits on the coupon, more than twice, than the AFRL 03 and CMAS particles.
To isolate the effects of particle material and composition on deposition percentage, simulation was carried out with equi-sized CMAS and
AFRL 02 particles with the particle diameter set to $20.15\ \mu m$. The cumulative deposition percentage along with the
cumulative average impact and rebound velocity is shown in Fig. \ref{fig:dep_vel_reb_dpeq}. Compared to the CMAS
particles, the AFRL 02 particles begin depositing on the coupon at a later time and have a lower percentage of
deposition at about 2.5\%. The average impact and rebound velocities in the normal direction are are quite comparable,
\begin{figure}[t]
\centering
  \begin{subfigure}
    \centering
    \includegraphics[scale=0.30]{images/afrl02_cmas_dpeq_sigY_dep_pct_shd}
  \end{subfigure}
  \begin{subfigure}
    \centering
    \includegraphics[scale=0.30]{images/afrl02_cmas_dpeq_sigY_Vimp_Vreb_shd}
  \end{subfigure}
  \caption{PLOT OF THE CUMULATIVE PERCENTAGE OF PARTICLES BEING DEPOSITED ON THE COUPON (TOP) AND THE AVERAGE IMPACT AND REBOUND
  VELOCITIES OF THE PARTICLES IN THE NORMAL DIRECTION (BOTTOM).}
\label{fig:dep_vel_reb_dpeq} 
\end{figure}
\section*{SUMMARY}

In this work, two-way coupled Euler-Lagrange simulations are carried  out to investigate sand particle deposition on a YSZ coupon using the deposition model proposed by Bons \et \cite{bons2017simple}. These simulations were modeled after the experiments performed at the HPIR facility
at the US Army Research Laboratory. Monodisperse particles representing two different materials, CMAS and AFRL sand were
used in this work. The CMAS were shown to have the highest amount of deposition of about 4\% on the coupon
followed by the smaller AFRL 02 sand particles with $\sim 3.5\%$. Notwithstanding the differences between the
experiments and the simulations in terms of the particle size distributions, the deposition rates were observed to be
quite comparable to the experiments. The larger AFRL 03 particles shown negligible amount of deposition amongst all the
different particles. The Stokes number of the AFRL 03 particles, although being the same order of magnitude as the rest,
could explain the ballistic behavior of these particles. Another interesting observation can be seen between the AFRL 03
particles and the CMAS particles in regards to the total number of impacts and the deposition percentages. Although both
these particles register a comparable number of impacts on the coupon, the amount deposited is starkly different. This
could possibly arise due to differences 1) in material properties or/and 2) in particle size. The results
from equi-sized CMAS and AFRL particles, with comparable $Stk$ 1.93 and 1.83 respectively, suggest that the differences
could be due to the material properties rather than the particle size. In addition to this, there are also uncertainties
associated with experimental data of the rebounding particles. Future work is reserved for investigations pertaining to
polydisperse particle mixtures and different TBC materials. An experimental campaign is underway at ARL which is aimed
at quantifying parameter uncertainty related to particle measurements used in the modeling. 

\bibliographystyle{asmems4}

\begin{acknowledgment}
  The authors acknowledge the support received by the Army Research Office Mathematical Sciences Division for this
  research. The views and conclusions contained in this document are those of the authors and should not be interpreted
  as representing the official policies or positions, either expressed or implied, of the U.S. Army Research Laboratory
  or the U.S. Government. The U.S. Government is authorized to reproduce and distribute reprints for Government purposes
  notwithstanding any copyright notation herein.

  This material is also based upon work supported by, or in part by, the Department of Defense (DoD) High Performance
  Computing Modernization Program (HPCMP) under User Productivity Enhancement, Technology Transfer, and Training (PET)
  contract \#47QFSA18K0111, TO\# ID04180146. We acknowledge Mr. Spencer Starr for his work on the implementation and
  validation of the deposition model in Athena-RFX.

  We gratefully acknowledge the computing resources provided on “Onyx” High Performance Computing cluster operated by
  the Department of Defense High Performance Computing Modernization Program (HPCMP).
  
  We gratefully acknowledge Dr. Alexei Poludnenko from the University of Connecticut as well as Dr. Prashant Khare from the University of Cincinnati for insightful discussions on multiphase fluid dynamics modeling. 
\end{acknowledgment}

\bibliography{refs}

\end{document}

%% file: dfn.tex
\def \et {\emph{et al.}}

\def \ug { \textbf{u} }

\def \Fqs { \textbf{F}_{qs}}

\def \fqs { \textbf{f}_{qs}}

\def \gqs { g_{qs}}

\def \Tp {\mathrm{T}_p}

\def \vp {\mathbf{V}}
\def \xp {\mathbf{X}}
\def \rp {\rho_p}

\def \div { \nabla \cdot }

\def \taub { \boldsymbol{\tau} }

\def \corni {\text{CoR\textsubscript{in}}}
\def \corn {\text{CoR\textsubscript{n}}}

\newcommand{\pp}[2]{\dfrac{\partial #1}{\partial #2} }
\newcommand{\dd}[2]{\dfrac{\mathrm d #1}{\mathrm d \mathrm{#2}} }

